\documentclass[sigconf]{acmart}
\usepackage{graphicx}
\usepackage{amsmath}
\usepackage{amsthm}
\usepackage{amsfonts}
\usepackage{multirow}
\usepackage{color}
\usepackage{bm}
\usepackage{balance}
\AtBeginDocument{%
  }

\setcopyright{acmlicensed}
\copyrightyear{2025}
\acmYear{2025}
\acmDOI{10.1145/3696410.3714832}

\acmConference[WWW '25] {Proceedings of the ACM Web Conference 2025}{April 28--May 2, 2025}{Sydney, NSW, Australia.}
\acmBooktitle{Proceedings of the ACM Web Conference 2025 (WWW '25), April 28--May 2, 2025, Sydney, NSW, Australia}
\acmISBN{979-8-4007-1274-6/25/04}

\begin{document}

\title{Multimodal Graph-Based Variational Mixture of Experts Network for Zero-Shot Multimodal Information Extraction}
\renewcommand{\shorttitle}{Multimodal Graph-Based Variational Mixture of Experts Network \\for Zero-Shot Multimodal Information Extraction}

\author{Baohang Zhou}
\affiliation{%
\institution{College of Computer Science\\VCIP, DISSec, TMCC, TBI Center\\Nankai University}
\city{Tianjin}
\country{China}
}
\email{zhoubaohang@dbis.nankai.edu.cn}

\author{Ying Zhang}
\authornote{Corresponding author.}
\affiliation{%
\institution{College of Computer Science\\VCIP, DISSec, TMCC, TBI Center\\Nankai University}
\city{Tianjin}
\country{China}
}
\email{yingzhang@nankai.edu.cn}

\author{Yu Zhao}
\affiliation{%
\institution{College of Computer Science\\VCIP, DISSec, TMCC, TBI Center\\Nankai University}
\city{Tianjin}
\country{China}
}
\email{zhaoyu@dbis.nankai.edu.cn}

\author{Xuhui Sui}
\affiliation{%
\institution{College of Computer Science\\VCIP, DISSec, TMCC, TBI Center\\Nankai University}
\city{Tianjin}
\country{China}
}
\email{suixuhui@dbis.nankai.edu.cn}

\author{Xiaojie Yuan}
\affiliation{%
\institution{College of Computer Science\\VCIP, DISSec, TMCC, TBI Center\\Nankai University}
\city{Tianjin}
\country{China}
}
\email{yuanxj@nankai.edu.cn}
\renewcommand{\shortauthors}{Baohang Zhou, Ying Zhang, Yu Zhao, Xuhui Sui, \& Xiaojie Yuan}
\newcommand{\figref}[1]{Figure \ref{#1}}
\newcommand{\tbref}[1]{Table \ref{#1}}
\renewcommand{\eqref}[1]{Eqn. \ref{#1}}

\begin{abstract}
  Multimodal information extraction on social media is a series of fundamental
  tasks to construct the multimodal knowledge graph.
  The tasks aim to extract the structural information in free texts with the
  incorporate images, including: multimodal named entity typing and multimodal
  relation extraction.
  However, the growing number of multimodal data implies a growing category set
  and the newly emerged entity types or relations should be recognized without
  additional training.
  To address the aforementioned challenges, we focus on the zero-shot multimodal
  information extraction tasks which require using textual and visual modalities
  for recognizing unseen categories.
  Compared with text-based zero-shot information extraction models, the existing
  multimodal ones make the textual and visual modalities aligned directly and
  exploit various fusion strategies to improve their performances.
  But the existing methods ignore the fine-grained semantic correlation of
  text-image pairs and samples.
  Therefore, we propose the multimodal graph-based variational mixture of
  experts network (MG-VMoE) which takes the MoE network as the backbone and
  exploits it for aligning multimodal representations in a fine-grained way.
  Considering to learn informative representations of multimodal data, we design
  each expert network as a variational information bottleneck to process two
  modalities in a uni-backbone.
  Moreover, we also propose the multimodal graph-based virtual adversarial
  training to learn the semantic correlation between the samples.
  The experimental results on the two benchmark datasets demonstrate the
  superiority of MG-VMoE over the baselines.
\end{abstract}

\ccsdesc[500]{Information systems~Multimedia and multimodal retrieval}
\ccsdesc[500]{Computing methodologies~Information extraction}

\keywords{Multimodal information extraction, Zero-shot learning, Multimodal representation learning}


\maketitle

\section{Introduction}
Extracting structural information from free text in conjunction with images, in
order to construct a multimodal knowledge
graph~\cite{9961954,DBLP:conf/sigir/ZhaoZZQSC24}, constitutes a series of
fundamental tasks known as multimodal information extraction (MIE).
The MIE tasks are associated with the entity information to complete the
specific tasks including: multimodal named entity typing
(MET)~\cite{zhang-etal-2023-incorporating-object} and multimodal relation
extraction (MRE)~\cite{DBLP:conf/mm/ZhengFFCL021}.
Compared with the text-based IE models, the multimodal-based ones are proposed
to capture the correlations of textual and visual contents with various fusion
strategies for effective entity and relation
classification~\cite{zhou-etal-2024-mcil}.

In practical situations, however, the number of intricate entity types and
relations is continually expanding, necessitating additional human input for
annotating every novel category that arises.
To address the above issue, the introduction of zero-shot learning into
text-based information extraction (ZS-IE) models facilitated the identification
of novel categories of entity types or relations without requiring additional
training.
The existing ZS-IE approaches primarily concentrate on the textual modality,
leveraging pre-trained language models to extract entity features for
constructing representations of type or relation prototypes.
\citet{DBLP:conf/coling/MaCG16} and \citet{chen-li-2021-zs} respectively
considered the names of types and relations as the prototypical knowledge for
recognizing the novel categories.
Moreover, the attention mechanism is applied on the ZS-IE models to extract the
fine-grained context representations implied in the external
descriptions~\cite{DBLP:conf/naacl/ObeidatFST19,DBLP:conf/acl/ZhaoZZZGWWPS23}.
With the development of social media, the growing number of multimodal data
implies expanding category set of types and relations.
And the above methods are focused on the textual modality while ignoring the
visual modality.

The vital challenge to exploit multimodal data is to bridge the semantic gap
between the two modalities.
The previous MIE models proposed the different fusion strategies or alignment
modules to extract the useful multimodal representations.
\citet{DBLP:conf/mm/ZhengFFCL021} designed the dual graph-based multimdoal
alignment and fusion modules to improve the MRE performance.
\citet{zhang-etal-2023-incorporating-object} exploited the cross-modal
transformer to obtain the multimodal representations for modeling the MET task.
\begin{figure}[t]
    \centering
    \includegraphics[width=0.45\textwidth]{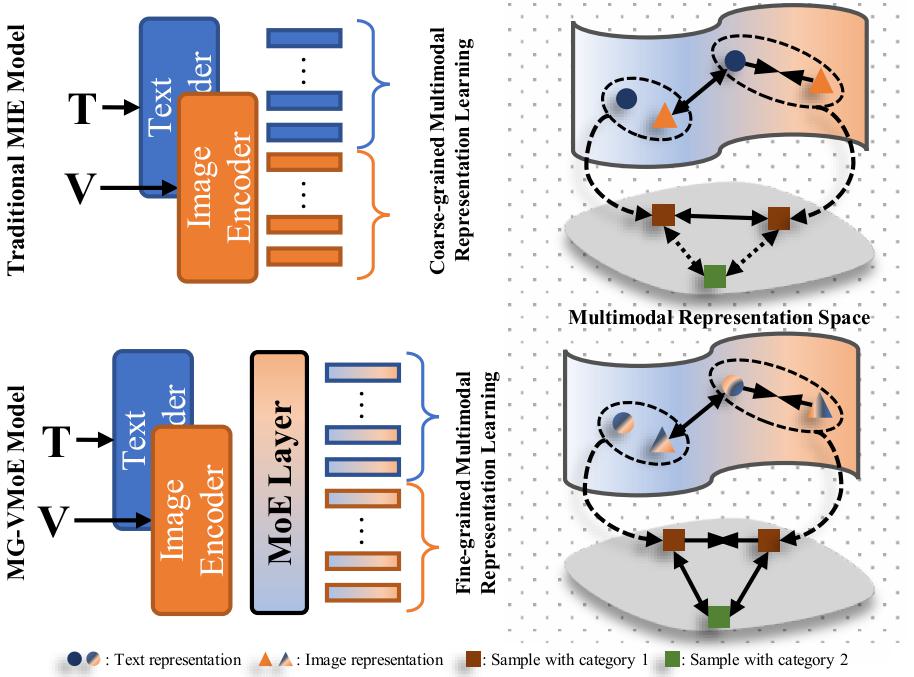}
    \setlength{\abovecaptionskip}{0.1cm}
    \caption{
        The multimodal representation space comparison between the traditional
        MIE and the MG-VMoE models.
    }
    \label{fig:introduction}
    \vspace{-13pt}
\end{figure}
However, the current MIE models lack efficiency in recognizing newly emerged
entity types or relationships on social media without additional training.  This
is due to the challenge posed by the diversity of textual and visual contents,
which span a wide range of entities and for which current models are unable to
effectively bridge the semantic gap between the two modalities. As illustrated
in \figref{fig:introduction}, traditional MIE models rely on coarse-grained
multimodal representation learning to align the global features of text-image
pairs within samples.  Since samples within the same category can exhibit
significant semantic variation in their texts and images, this approach is
insufficient for capturing fine-grained semantic correlations between the two
modalities at the token level and for clustering multimodal samples of the same
category which constraints its ability to establish connections between
multimodal samples and prototypical categories.

To address the above limitations, we propose the \textbf{m}ultimodal
\textbf{g}raph-based \textbf{v}ariational \textbf{m}ixture \textbf{o}f
\textbf{e}xperts (MG-VMoE)~\footnote{\url{https://github.com/ZovanZhou/MG-VMoE}}
network to tackle the zero-shot MIE (ZS-MIE) task.
The MG-VMoE network is based on the fine-grained multimodal representation
learning which consists of the architectures and the specific training method.
For capturing semantic correlations between the two modalities at the token
level, we utilize the mixture of experts (MoE) network as the backbone which
exploits the sparse expert weights to align the textual and visual token
representations in a fine-grained way.
With the purpose to model the informative representations of multimodal data, we
design each expert network as a variational information bottleneck (VIB) to
handle the two modalities in a uni-backbone.
For clustering samples belonging to the same category, we propose a multimodal
graph-based virtual adversarial training method to capture the semantic
correlations between multimodal samples.
Ultimately, we fuse textual entity representations with multimodal ones through
an attention layer and measure the semantic similarity between the fusion
features and prototypical ones of different categories for recognition.
The contributions of this manuscript can be summarized as follows:
\begin{itemize}
    \item We present the zero-shot multimodal information extraction (ZS-MIE)
    task which leverages the text and image pairs to extract the novel knowledge
    such as: entity types or relations on social media without additional
    training.
    \item We propose a multimodal graph-based variational mixture of experts
    (MG-VMoE) network based on the fine-grained multimodal representation
    learning.
    Not only does the network utilize the VMoE architecture to model the aligned
    multimodal representations within individual samples, but it also leverages
    multimodal graph-based virtual adversarial training to capture semantic
    correlations between samples.
    \item We conduct the extensive experiments on the two benchmark MIE datasets
    and the experimental results demonstrate the superiority of the proposed
    model over baselines.
\end{itemize}

\section{Related Work}

\subsection{Zero-shot Information Extraction}
Information extraction (IE) encompasses a range of tasks, notably named entity
typing and relation extraction, aimed at distilling structural information from
unstructured texts for the purpose of constructing comprehensive knowledge
graphs~\cite{DBLP:journals/corr/abs-2302-05019}.
Considering to recognize the unseen categories like: entity types or relations
without additional training, the zero-shot learning was introduced into
traditional information extraction (ZS-IE) tasks.
The vital challenge for ZS-IE is to learn generalizable representations of
entities and prototypical knowledge of categories.

For zero-shot named entity typing (ZS-ET), \citet{DBLP:conf/coling/MaCG16}
firstly proposed a label embedding method to encode the prototypical knowledge
of types with textual embeddings, and bridge the semantic correlation between
entity mentions and types.
\citet{DBLP:conf/www/RenLZ20} employed the attention mechanism to extract local
features that are relevant to the types, with a focus on the nuanced semantic
representations of both mentions and their contexts.
\citet{DBLP:conf/coling/ZhangXLY20} devised the ZS-ET model, augmented with
memory capabilities, to retain observed types as memory elements and facilitate
knowledge transfer from known to unknown types, thereby explicitly capturing the
semantic relationship between them.
Furthermore, auxiliary data including descriptions sourced from websites was
integrated into the ZS-ET model to augment the representation of mentions and
types~\cite{DBLP:conf/naacl/ObeidatFST19,DBLP:conf/emnlp/0019JL0FYX21}.
For zero-shot relation extraction (ZS-RE), \citet{chen-li-2021-zs} initially
leveraged BERT to acquire two functions, which project entities and relation
descriptions into an embedding space by concurrently minimizing the distances
between them and subsequently categorizing their corresponding relations.
\citet{DBLP:conf/acl/ZhaoZZZGWWPS23} introduced a fine-grained semantic matching
method, which dissects the overall sentence-level similarity score into distinct
components for entity and context matching.
\citet{DBLP:conf/coling/GongE24a} presented a prompt-driven model that augments
semantic knowledge by creating instances featuring unseen relations from
existing instances with known relations.

In essence, the current ZS-IE approaches solely concentrate on textual modality,
neglecting the potential semantic enrichment from visual content. In contrast to
these endeavors, our focus lies in the ZS-MIE task, aimed at extracting
innovative structural knowledge embedded within multimodal data.

\subsection{Mulitmodal Information Extraction}
As the volume of multimodal data continues to expand, researchers have extended
the traditional IE tasks to encompass multimodal IE, resulting in improved
outcomes.
\citet{DBLP:conf/naacl/MoonNC18} initially broadened the scope of traditional
text-based named entity recognition to encompass multimodal named entity
recognition (MER), introducing the modality attention module to integrate
textual information with image one, thereby enhancing the accuracy of sequence
label predictions.
\citet{zhang-etal-2023-incorporating-object} proposed to incorporate visual
objects and exploit the cross-modal transformer to obtain multimodal
representations for tackling the multimodal named entity typing (MET) task
firstly.
\citet{DBLP:conf/mm/ZhengFFCL021} introduced the multimodal relation extraction
(MRE) task, leveraging visual modality to bolster the semantic representations
of textual modality.
\citet{DBLP:journals/taslp/CuiCCSLLS24} exploited the variational information
bottleneck to extract effective multimodal representations for the MIE tasks.

In summary, the aforementioned multimodal information extraction tasks operate
within a supervised framework. However, our focus lies in zero-shot multimodal
information extraction (ZS-MIE). In contrast to supervised learning that relies
on abundant labeled data, zero-shot learning prioritizes the development of
generalizable representations for both samples and semantic labels, enabling the
inference of samples belonging to unobserved categories.

\section{Preliminary}
Before introducing the details of the proposed model, we formalize the problem of zero-shot multimodal information extraction (ZS-MIE).
We obtain the training dataset $\mathcal{D}_{train}$ with the seen category set $\mathcal{Y}_{s}$ and the test dataset $\mathcal{D}_{test}$ with the unseen category set $\mathcal{Y}_{u}$.
The seen category set is defined as $\mathcal{Y}_{s} = \{y_1^s, y_2^s, \dots, y_{|\mathcal{Y}_s|}^s\}$ and the unseen one is denoted as $\mathcal{Y}_{u} = \{y_1^u, y_2^u, \dots, y_{|\mathcal{Y}_u|}^u\}$.
And the two aforementioned category sets are mutually disjoint.
Each sample of the datasets is denoted as a tuple $S = (T,V,E,Y)$ where $T$ denotes a natural language sentence, $V$ is the image coupled with the sentence, $E$ represents the entity information such as: the location in the sentence, and $Y$ is the label for specific tasks.
According to zero-shot learning, we need to use the training set
$\mathcal{D}_{train}$ to train a ZS-MIE model $\mathcal{M}$, i.e.,
$\mathcal{M}(S) \to Y \in \mathcal{Y}_s$ and evaluate it by recognizing the
unseen categories on the test set.

\section{Methodology}
\begin{figure*}[t]
    \centering
    \includegraphics[width=\textwidth]{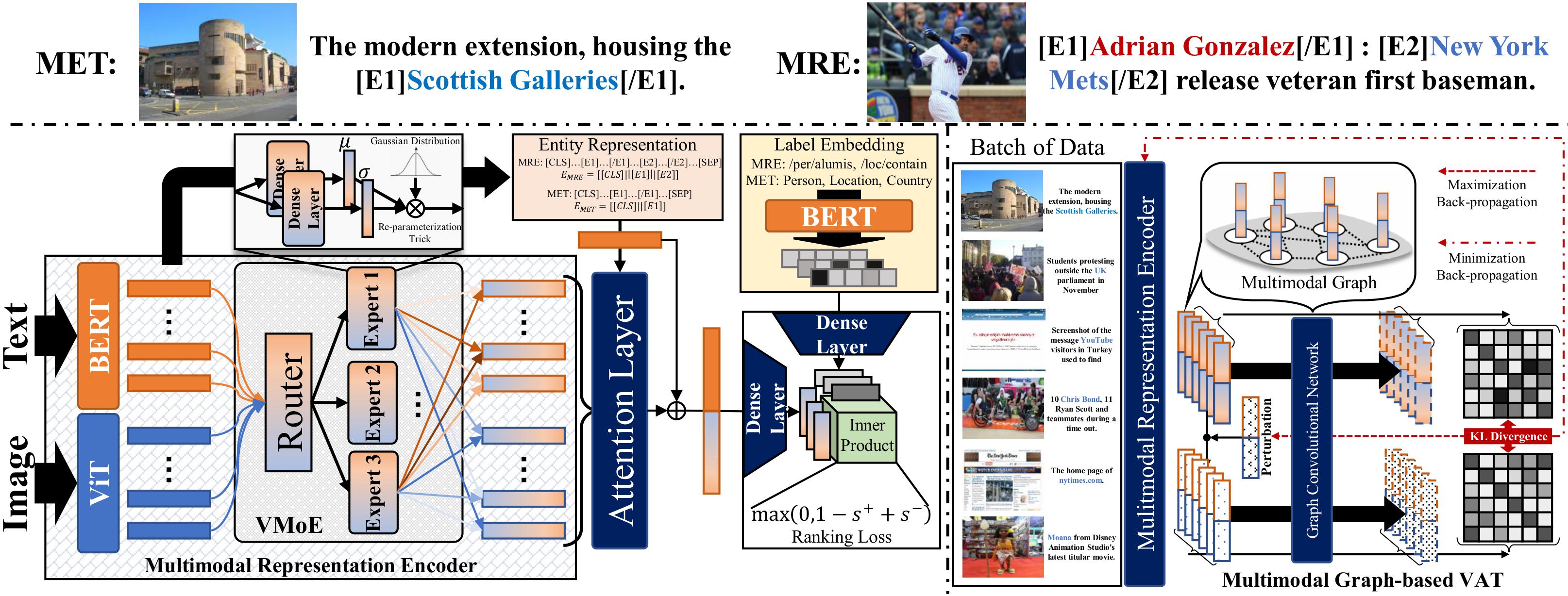}
    \setlength{\abovecaptionskip}{-0.3cm}
    \caption{
        The overall framework of the multimodal graph-based variational mixture of experts (MG-VMoE) network for zero-shot multimodal information extraction.
        The upper part is the samples of multimodal named entity typing and multimodal relation extraction.
        The lower left is the multimodal backbone network based on VMoE and the lower right is the multimodal graph-based vitrual adversarial training.
    }
    \label{fig:model}
    \vspace{-12pt}
\end{figure*}

In this section, we introduce the multimodal graph-based variational mixture of
experts (MG-VMoE) network for zero-shot multimodal information extraction as
shown in \figref{fig:model}.
The overall framework is based on the fine-grained multimodal representation
learning which consists of the architectures and specific training method.
The details can be summarized as the following parts:
(1) Firstly, we extract the multimodal input representations of the samples with
the pre-trained language and vision models.
(2) Secondly, to capture the fine-grained semantic correlation between the text
and image token representations, we propose the VMoE network which models the
informative and aligned representations of multimodal data in a uniform
backbone.
(3) Thirdly, we propose the multimodal-graph based virtual adversarial training
to model the semantic correlation between multimdoal samples and keep samples
belonging to the same category more clustered.
(4) Eventually, to identify unseen categories, we calculate distances between
the multimodal features of samples and label semantic ones.

\subsection{Multimodal Input Representation}
Given the multimodal samples which consists of texts and images, we need to map
them into the dense representations for deep neural networks.
For the images, we use ViT~\cite{DBLP:conf/iclr/DosovitskiyB0WZ21} to extract
the visual representations of images.
%
%
We input the image $V$ of the sample into ViT and obtain the visual
representations which are denoted as $\textbf{V} =
\{\textbf{v}_1,\textbf{v}_2,\dots,\textbf{v}_{|V|}\}$ where $\textbf{v}_i \in
\mathbb{R}^d$ and $|V|$ is the number of feature vectors output from ViT.

As for the texts, we denote the original sentence with $|S|$ words as $S=\{w_1,w_2,\dots,w_{|S|}\}$.
To take advantage of pre-trained language models, we utilize BERT~\cite{DBLP:conf/naacl/DevlinCLT19} as the text encoder to map discrete words into the dense representations.
Before feeding the sentence into BERT, we should pre-process it with special tokens.
Each sentence requires the insertion of the reserved tokens \texttt{[CLS]} at the beginning and \texttt{[SEP]} at the end.
Furthermore, the entity information holds significant importance for ZS-MIE tasks, which encompass MET and MRE.
The reserved tokens \texttt{[E1]}, \texttt{[/E1]} (and \texttt{[E2]}, \texttt{[/E2]}) are inserted into the sentence to mark the begin and end of the entities from the entity set $E$~\cite{zhou-etal-2024-mcil}.
Formally, the extended sentence with the special tokens is denoted as $\tilde{S}$.
And the calculation process of sentence representations can be simplified as $\textbf{T} = BERT(\tilde{S})$ where $\textbf{T} = \{\textbf{t}_{\texttt{[CLS]}}, \textbf{t}_1, \dots, \textbf{t}_{\texttt{[SEP]}}\} \in \mathbb{R}^{d \times |T|}$ and $|T|$ is the token number of the extended sentence.
We extract the entity representation according to the specific tasks.

\noindent \textbf{Multimodal Named Entity Typing (MET).}
To keep the entity and contextual information for MET task, we define the entity representation as:
$
    \textbf{E} = \left[\textbf{t}_{\texttt{[CLS]}} \oplus \textbf{t}_{\texttt{[E1]}}\right] \in \mathbb{R}^{2d}
$
where $\textbf{t}_{\texttt{[CLS]}}$ and $\textbf{t}_{\texttt{[E1]}}$ represent the features of the tokens \texttt{[CLS]} and \texttt{[E1]} respectively, and $\oplus$ is the vector concatenation operation.

\noindent \textbf{Multimodal Relation Extraction (MRE).}
Analogous to the MET task, we require extracting the entity representation by incorporating both the head and tail entities as:
$
    \textbf{E} = \left[\textbf{t}_{\texttt{[CLS]}} \oplus \textbf{t}_{\texttt{[E1]}} \oplus \textbf{t}_{\texttt{[E2]}} \right] \in \mathbb{R}^{3d}
$
where $\textbf{t}_{\texttt{[E1]}}$ and $\textbf{t}_{\texttt{[E2]}}$ denote the beginning tokens \texttt{[E1]} and \texttt{[E2]} of head and tail entities respectively.

Furthermore, we consider the semantic information of categories such as: the names of labels as the prototypical knowledge.
Given the seen category set $\mathcal{Y}_{s} = \{y_1^s, y_2^s, \dots, y_{|\mathcal{Y}_s|}^s\}$, we consider each category name as a sentence $\{w_1,w_2,\dots,w_l\}$ and feed it into BERT to acquire the textual representations $\{\textbf{r}_{0},\textbf{r}_1,\dots,\textbf{r}_{l+1}\}$.
The semantic representation of the category is calculated as $\textbf{R}=\frac{1}{l+2}\sum_{i=0}^{l+1}\textbf{r}_i$.
Therefore, the prototypical representations of the seen category set is defined as $\textbf{C}_s = \{\textbf{R}_1^s,\textbf{R}_2^s,\dots,\textbf{R}_{|\mathcal{Y}_s|}^s\}$.

\subsection{Variational Mixture of Experts Network}
The multimodal representations from the pre-trained language and vision models exist in their respective modality spaces.
To address the semantic discrepancy between the two modalities, the existing MIE models frequently resort to contrastive learning to harmonize the multimodal representations emanating from pre-trained language and vision models~\cite{DBLP:conf/wsdm/XuHSW22}.
Nonetheless, the aforementioned coarse-grained approach to multimodal representation learning primarily emphasizes the holistic representations of multimodal data, neglecting to model the local semantic relationships between text and image token representations.
Therefore, we propose the variational mixture of experts (VMoE) network as the backbone to make the multimodal data aligned uniformly.
The traditional MoE network consists of the router module and the expert modules~\cite{DBLP:conf/nips/MustafaRPJH22}.
Upon inputting a feature vector into the MoE network, the router module initially determines which expert modules to activate, guided by the input data's characteristics. Subsequently, each activated expert module processes the input data individually, producing respective outputs. Ultimately, these outputs undergo a weighted summation process, with weights assigned by the router module, to yield the fused prediction result~\cite{DBLP:conf/nips/BaoW0LMASPW22}.
Compared with the traditional MoE network, we formalize each expert module as the variational information bottleneck (VIB)~\cite{DBLP:journals/taslp/CuiCCSLLS24} which can keep the multimodal representations from VMoE informative and aligned.

Given the textual representation $\textbf{T}$ and the visual one $\textbf{V}$, we combine them with the direct concatenation as $\textbf{M} = \left[\textbf{V};\textbf{T}\right] \in \mathbb{R}^{d \times (|V|+|T|)}$.
The expert module, for multimodal representation $\textbf{M}$, is structured as VIB that learns a latent representation $\textbf{Z}$ while preserving sufficient information from $\textbf{M}$ crucial for prediction.
The information bottleneck (IB)~\cite{DBLP:conf/itw/TishbyZ15} is formalized as follows:
\begin{equation}
    \mathcal{L}_{IB} = \beta \cdot I(\textbf{M},\textbf{Z}) - I(\textbf{Z}, Y)
\end{equation}
where $I(\cdot)$ is the mutual information (MI) between two variables.
To reduce irrelevant information, we minimize the mutual information $I(\textbf{M},\textbf{Z})$ between the input representation $\textbf{M}$ and the latent representation $\textbf{Z}$ generated by the expert module. Additionally, to ensure sufficient information for prediction, we maximize the mutual information $I(\textbf{Z}, Y)$ between the latent representation $\textbf{Z}$ and the target label $Y$.
Considering that the MI is computationally intractable for deep neural networks~\cite{DBLP:journals/taslp/CuiCCSLLS24}, we utilize the variational manner to encode the latent representation.
Therefore, the latent gaussian distributional variable $\textbf{Z}$ is defined as follows:
\begin{align}
    \textbf{Z} \sim \mathcal{N}(\bm{\mu}, \bm{\sigma}^2), \quad \bm{\mu} = FFNN(\textbf{M};\theta_{\mu}),\quad \bm{\sigma} = FFNN(\textbf{M};\theta_{\sigma})
\end{align}
where $\bm{\mu}$ and $\bm{\sigma}$ are the mean and standard deviation vectors, and $FFNN(\cdot;\theta)$ is short for the feed-forward neural network with the trainable parameter $\theta$.
We use the re-parameterization trick~\cite{DBLP:journals/corr/KingmaW13} to perform the equivalent sampling to obtain the latent representation $\textbf{Z}$ as the following equation:
\begin{equation}
    \textbf{Z} = \bm{\mu} + \bm{\sigma} \odot \epsilon, \quad \epsilon \sim \mathcal{N}(0, \textbf{I})
\end{equation}
where $\odot$ is the element-wise production and $\textbf{Z} \in \mathbb{R}^{d \times (|V|+|T|)}$ is the representation from each expert module.
And for the $K$ expert modules, the combination of their output latent representations is denoted as: $\{\textbf{Z}_i | \textbf{Z}_i \sim \mathcal{N}(\bm{\mu}_i, \bm{\sigma}_i^2), i=1,2,\dots,K\}$.
To fuse the representations from different expert modules, the router module predicts the gating weights corresponding to the $K$ expert modules.
The router module is implemented with the dense connection layer, and the gating weights are calculated as $\textbf{G} = softmax(\textbf{W}_g^T \textbf{M} + \textbf{b}_g) \in \mathbb{R}^{K \times (|V|+|T|)}$ where $\textbf{W}_g \in \mathbb{R}^{d \times K}$ and $\textbf{b}_g \in \mathbb{R}^{K}$ are the trainable parameters.
The fusion multimodal representations are calculated as: $\textbf{H} = \sum_{i=1}^K \textbf{Z}_i \cdot \textbf{G}_i$ and the fusion textual and visual token features are $\tilde{\textbf{V}} = \{\textbf{h}_i\}_{i=1}^{|V|}$ and $\tilde{\textbf{T}} = \{\textbf{h}_i\}_{i=|V|+1}^{|V|+|T|}$.
Considering to sparsely activate expert modules for each token representation, we exploit the entropy auxiliary loss to optimize the router module as the following equation:
\begin{equation} \label{eqn:loss-auxiliary}
    \mathcal{L}_{aux} = - \frac{1}{|V|+|T|} \sum_{j=1}^{|V|+|T|} \sum_{i=1}^{K} G_{i,j} \log G_{i,j}
\end{equation}
where $G_{i,j}$ is the weight score of the $i$-th expert module to the $j$-th token representation.

To optimize the model by the IB principle, we can estimate $I(\textbf{M}, \textbf{Z})$ as follows:
\begin{align}
    I(\textbf{M}, \textbf{Z}) & = KL(p(\textbf{Z}|\textbf{M}) \parallel p(\textbf{Z})) \leq KL(q(\textbf{Z}|\textbf{M})\parallel p(\textbf{Z}))
\end{align}
where the posterior distribution $p(\textbf{Z}|\textbf{M})$ could be approximated by the variational posterior distribution $q(\textbf{Z}|\textbf{M})$ and the prior distribution $p(\textbf{Z})$ is assumed as normal Gaussian distribution.
Therefore, the regularization loss of VIB for the latent representations from $K$ expert modules is defined as follows:
\begin{equation} \label{eqn:loss-regularization}
    \mathcal{L}_{reg} = \sum_{i=1}^K KL(q(\textbf{Z}_i|\textbf{M}) \parallel p(\textbf{Z}_i)) = \sum_{i=1}^K KL(\mathcal{N}(\bm{\mu}_i, \bm{\sigma}_i^2) \parallel \mathcal{N}(0, \textbf{I}))
\end{equation}
To keep the latent representations related to task labels, we maximize $I(\textbf{Z}, Y)$ with the variational lower bound~\cite{DBLP:journals/taslp/CuiCCSLLS24} of it as follows:
\begin{equation} \label{eqn:objective-function}
    I(\textbf{Z}, Y) = \mathbb{E}_{p(\textbf{Z}|\textbf{M})} \left[\log p(Y|\textbf{M})\right] \ge \mathbb{E}_{q(\textbf{Z}|\textbf{M})} \left[\log p(Y|\textbf{M})\right].
\end{equation}
The above equation could be optimized by the task loss function such as ranking loss function for ZS-MIE tasks. 

\subsection{Multimodal Graph-based Virtual Adversarial Training}
To bridge the semantic gap between the textual and visual modalities, the multimodal methods~\cite{DBLP:conf/wsdm/XuHSW22,DBLP:conf/nips/BaoW0LMASPW22} always utilize the contrastive learning to align the text and image representations of a sample.
Given a batch of $N$ samples $\{(T_i,V_i)\}_{i=1}^N$, we feed them into the multimodal representation encoder which includes the pre-trained language and vision models stacked with VMoE network to obtain the fusion multimodal representations as $\{\textbf{H}_i|\textbf{H}_i=\left[\tilde{\textbf{V}}_i;\tilde{\textbf{T}}_i\right],i=1,2,\dots,N\}$.
We average the fusion textual and visual token features as the global representations $\{(\bar{\textbf{T}}_i, \bar{\textbf{V}}_i)|\bar{\textbf{T}}_i=\frac{1}{|T|}\sum_{j=1}^{|T|}\tilde{\textbf{t}}_{i,j}, \bar{\textbf{V}}_i=\frac{1}{|V|}\sum_{j=1}^{|V|}\tilde{\textbf{v}}_{i,j},i=1,2,\dots,N\}$.
The objective of multimodal contrastive learning is to discern matched pairs from among $N \times N$ potential image-text combinations, ensuring that representations of paired inputs are positioned closer together in the representation space compared to those of unpaired inputs.
Therefore, the multimodal contrastive learning for one batch is defined as follows:
\begin{equation} \label{eqn:loss-contrastive-learning}
    \mathcal{L}_{cl} = \sum_{i=1}^{N}- \frac{1}{2} (\log \frac{\exp(\bar{\textbf{T}}_i^T \bar{\textbf{V}}_i)}{ \sum_{j=1}^N \exp(\bar{\textbf{T}}_i^T \bar{\textbf{V}}_j)} + \log \frac{\exp(\bar{\textbf{T}}_i^T \bar{\textbf{V}}_i)}{ \sum_{j=1}^N \exp(\bar{\textbf{T}}_j^T \bar{\textbf{V}}_i)} )
\end{equation}

However, the aforementioned coarse-grained contrastive learning approach solely emphasizes the semantic coherence between text-image pairs within individual samples, falling short in clustering samples of the same category while simultaneously discerning intricate semantic nuances within multimodal data. This limitation ultimately restricts its capacity to forge fine-grained correlation between multimodal samples and prototypical knowledge.
Therefore, we propose the multimodal-graph based virtual adversarial training (MG-VAT) to model the semantic correlation between samples.
Given a batch of multimodal samples, we integrate the global textual and visual representations as a whole $\textbf{P} = \{\bar{\textbf{H}}_i|\bar{\textbf{H}}_i=\left[\bar{\textbf{V}}_i\oplus\bar{\textbf{T}}_i\right],i=1,2,\dots,N\} \in \mathbb{R}^{2d \times N}$.
To measure the semantic similarities of samples, we construct the multimodal sample correlation graph $\textbf{A} = \{a_{i,j}|i,j\in\{1,2,\dots,N\}\} \in \mathbb{R}^{N \times N}$.
And the elements in the graph are calculated as follows:
\begin{equation}
    a_{i,j} = a_{j,i} = ( 1 + \frac{\bar{\textbf{H}}_i^T\bar{\textbf{H}}_j}{\Vert \bar{\textbf{H}}_i \Vert_2 \Vert \bar{\textbf{H}}_j \Vert_2} ) / 2.
\end{equation}
We utilize the graph information to aggregate the multimodal representations of relevant samples as: $\hat{\textbf{P}} = \{\hat{\textbf{H}}_i\}_{i=1}^N = \textbf{P}\textbf{A} $ and the irrelevant ones as: $\hat{\textbf{P}}^\prime = \{\hat{\textbf{H}}_i^\prime\}_{i=1}^N = \textbf{P}(\textbf{1} - \textbf{A})$.
For each sample, we can measure its relative semantic correlation score with other relevant and irrelevant samples in the batch as:
\begin{equation}
    \textbf{s}_i = softmax([\bar{\textbf{H}}_i^T\hat{\textbf{H}}_i;\bar{\textbf{H}}_i^T\hat{\textbf{H}}_i^\prime]) \in \mathbb{R}^2
\end{equation}
and the score of the batch is denoted as $\textbf{S} = \{\textbf{s}_i\}_{i=1}^N \in \mathbb{R}^{2 \times N}$.

To enhance the robustness of multimodal representations of samples, we employ VAT~\cite{DBLP:journals/corr/GoodfellowSS14,DBLP:journals/pami/MiyatoMKI19} to minimize the KL divergence between the relative semantic correlation scores of original samples and those of adversarial samples based on the multimodal sample correlation graph.
We introduce the perturbation vector $\bm{\tau} \in \mathbb{R}^{2d}$ to generate the multimodal representations of adversarial samples as $\tilde{\textbf{P}} = \{\tilde{\textbf{H}}_i|\tilde{\textbf{H}}_i=\bar{\textbf{H}}_i + \bm{\tau},i=1,2,\dots,N\}$.
Based on the given graph $\textbf{A}$, we can also calculate the relative semantic correlation scores of adversarial samples as $\tilde{\textbf{S}}$.
In order to reduce the influence of perturbation, the adversarial loss for VAT is defined as follows:
\begin{equation} \label{eqn:loss-vat}
    \mathcal{L}_{vat} = KL(p(\textbf{S}|\textbf{P}) \parallel p(\tilde{\textbf{S}}|\tilde{\textbf{P}})).
\end{equation}
Moreover, to compute the worst perturbation which can significantly improve multimodal representations, we can optimize the perturbation $\bm{\tau}$ by the following objective function:
\begin{equation} \label{eqn:loss-perturbation}
    \mathop{\arg \max}_{\bm{\tau}} KL(p(\textbf{S}|\textbf{P}) \parallel p(\tilde{\textbf{S}}|\tilde{\textbf{P}})) - \Vert \bm{\tau} \Vert_2.
\end{equation}

\subsection{Training and Inference Procedure}
Given the fusion multimodal representations $\textbf{H}$ and the textual entity representation $\textbf{E}$, we utilize the attention mechanism~\cite{DBLP:conf/acl/ZhouSTQLHX16} to extract the local features of the former which are related to the latter.
The attention score is defined as $\alpha_i = \frac{\exp(\textbf{W}_a^T[\textbf{h}_i\oplus\textbf{E}] + \textbf{b}_a)}{\sum_{j=1}^{|V|+|T|}\exp(\textbf{W}_a^T[\textbf{h}_j\oplus\textbf{E}] + \textbf{b}_a)}$ where $\textbf{W}_a \in \mathbb{R}^{d+|\textbf{E}|}$ and $\textbf{b}_a \in \mathbb{R}$ are the trainable parameters and $|\textbf{E}|$ is the dimension number of textual entity representation.
Therefore, the entity-aware multimodal representation is calculated as $\textbf{U} = \sum_{i=1}^{|V|+|T|}\alpha_i\textbf{h}_i$.
For predicting the category, we concatenate the textual entity representation and entity-aware one as $\tilde{\textbf{U}} = \left[\textbf{U}\oplus\textbf{E}\right]$ and regard the the semantic representations $\textbf{C}_s = \{\textbf{R}_1^s,\textbf{R}_2^s,\dots,\textbf{R}_{|\mathcal{Y}_s|}^s\}$ of the seen category set as the prototypical knowledge.
To assess the association between the sample and its category, we employ the semantic similarity to determine the score using the formula as $o_i = \hat{\textbf{U}}^T\hat{\textbf{R}}_i$ where $o_i$ is the score between the sample and the $i$-th category in $\textbf{C}_s$.
And the textual entity representation and entity-aware one are projected into the shared semantic space as $\hat{\textbf{U}} = FFNN(\tilde{\textbf{U}};\theta_{o1}) \in \mathbb{R}^h$ and $\hat{\textbf{R}}_i = FFNN(\textbf{R}_i;\theta_{o2}) \in \mathbb{R}^h$.
Based on the above score, we leverage the ranking loss to ensure that the score of the true label remains higher than those of the false labels.
The objective of max-margin ranking is defined as follows:
\begin{equation} \label{eqn:loss-ranking}
    \mathcal{L}_{rank} = \sum_{i=1, y_i^s \neq Y}^{|Y_s|} \max(0, 1 - o^{+} + o_i)
\end{equation}
where $o^{+}$ is the score of the true type $y_i^s = Y$.
And the objective function of \eqref{eqn:objective-function} can also be optimized by the task-relevant loss.

Given the batches of multimodal samples, we feed them into the model and firstly update the perturbation vector by \eqref{eqn:loss-perturbation}.
To train the model with different objectives at once, we introduce the hyper-parameter to sum the corresponding losses.
The overall loss is defined as follows:
\begin{equation} \label{eqn:loss-overall}
    \mathcal{L} = \mathcal{L}_{rank} + \beta \cdot (\mathcal{L}_{aux} + \mathcal{L}_{reg} + \mathcal{L}_{cl} + \mathcal{L}_{vat})
\end{equation}
where $\beta$ is the hyper-parameter to balance the different losses.
Subsequently, we employ stochastic gradient descent (SGD) techniques to update model weights based on the loss calculated by \eqref{eqn:loss-overall}.
In the inference phase, we evaluate the scores between samples in $D_{test}$ and
unseen categories in $Y_u$, and designate the category scoring the highest as
the predicted outcome.

\section{Experiments}

\subsection{Datasets and Experiment Settings}
We delve into the realm of zero-shot multimodal information extraction, specifically focusing on two tasks: multimodal named entity typing (MET) and multimodal relation extraction, both executed within a zero-shot setting. For these tasks, we undertake experiments utilizing the respective benchmark datasets.
For MET task, we utilize the WikiDiverse~\cite{wang-etal-2022-wikidiverse} as the benchmark dataset.
Each sample in the dataset consists of a text-image pair sourced from Wikinews, with the entity mention in the sentence manually annotated into one of 13 fine-grained types.
For the MRE task, we utilize the benchmark dataset proposed by \citet{DBLP:conf/mm/ZhengFFCL021}, which is based on Twitter posts. Annotators randomly selected samples covering various topics. The MRE dataset contains samples categorized into 23 relation types.
Since the above two datasets include meaningless categories such as "Other" or "None", we exclude them and only consider categories with actual semantics.

\begin{table}[t]
  \setlength{\abovecaptionskip}{0.1cm}
  \caption{
      Performance comparison on the MET and MRE datasets under the zero-shot
      settings.
      The bold numbers indicate that the improvement of MG-VMoE over traditional baselines is statistically significant with $p < 0.01$ under t-test.
  }
  \label{tab:main-results}
  \resizebox{\columnwidth}{!}{
  \begin{tabular}{lcccc}
      \bottomrule
      \multicolumn{5}{c}{Multimodal Named Entity Typing} \\
      Model            & Precision            & Recall               & F1                   & Accuracy             \\ \hline
      Text             & \multicolumn{1}{l}{} & \multicolumn{1}{l}{} & \multicolumn{1}{l}{} & \multicolumn{1}{l}{} \\ \hline
      Proto            & 23.9 $\pm$ 5.9       & 24.6 $\pm$ 3.5       & 13.9 $\pm$ 5.7       & 29.3 $\pm$ 11.7      \\
      DBZFET           & 30.9 $\pm$ 4.9       & 29.5 $\pm$ 4.2       & 16.2 $\pm$ 1.3       & 23.6 $\pm$ 1.4       \\
      NZFET            & 26.3 $\pm$ 9.0       & 26.8 $\pm$ 4.8       & 13.3 $\pm$ 1.8       & 18.4 $\pm$ 2.6       \\
      MZET             & 29.9 $\pm$ 3.2       & 28.8 $\pm$ 5.6       & 11.4 $\pm$ 1.5       & 21.1 $\pm$ 6.5       \\ \hline
      Multimodal       & \multicolumn{1}{l}{} & \multicolumn{1}{l}{} & \multicolumn{1}{l}{} & \multicolumn{1}{l}{} \\ \hline
      MMProto          & 27.3 $\pm$ 3.8       & 27.2 $\pm$ 4.1       & 17.1 $\pm$ 9.2       & 24.4 $\pm$ 12.4      \\
      MOVCNet          & 28.6 $\pm$ 5.0       & 24.0 $\pm$ 3.3       & 13.9 $\pm$ 5.0       & 23.3 $\pm$ 10.8      \\
      LLaVA            & 40.9 $\pm$ 2.8       & 51.6 $\pm$ 6.4       & 39.5 $\pm$ 2.9       & 53.5 $\pm$ 9.7       \\
      Ours             & $\textbf{37.1} \pm 3.3$  & $\textbf{31.3} \pm 4.9$  & $\textbf{23.7} \pm 2.0$  & $\textbf{45.1} \pm 2.0$ \\ \bottomrule

      \bottomrule
      \multicolumn{5}{c}{Multimodal Relation Extraction} \\
      Model            & Precision            & Recall               & F1                   & Accuracy             \\ \hline
      Text             & \multicolumn{1}{l}{} & \multicolumn{1}{l}{} & \multicolumn{1}{l}{} & \multicolumn{1}{l}{} \\ \hline
      Proto            & 24.5 $\pm$ 12.5      & 27.2 $\pm$ 10.2      & 21.6 $\pm$ 9.6       & 38.0 $\pm$ 9.3       \\
      ZS-BERT          & 29.5 $\pm$ 4.8       & 29.3 $\pm$ 5.4       & 22.9 $\pm$ 4.1       & 37.2 $\pm$ 8.7       \\
      RE-Matching      & 28.0 $\pm$ 4.8       & 31.1 $\pm$ 7.1       & 22.7 $\pm$ 11.1      & 32.4 $\pm$ 17.4      \\
      ZS-SKA           & 29.9 $\pm$ 7.4       & 21.8 $\pm$ 6.0       & 16.6 $\pm$ 7.1       & 26.8 $\pm$ 11.3      \\ \hline
      Multimodal       & \multicolumn{1}{l}{} & \multicolumn{1}{l}{} & \multicolumn{1}{l}{} & \multicolumn{1}{l}{} \\ \hline
      MMProto          & 33.8 $\pm$ 3.9       & 29.8 $\pm$ 4.5       & 23.4 $\pm$ 4.8       & 35.6 $\pm$ 6.2      \\
      MOVCNet          & 27.6 $\pm$ 11.1      & 25.4 $\pm$ 7.4       & 19.8 $\pm$ 6.2       & 33.8 $\pm$ 10.5      \\
      LLaVA            & 25.1 $\pm$ 4.6       & 18.2 $\pm$ 2.0       & 13.6 $\pm$ 0.7       & 14.9 $\pm$ 3.1       \\
      Ours             & $\textbf{34.5} \pm 8.1$  & $\textbf{32.8} \pm 5.1$  & $\textbf{27.3} \pm 8.1$  & $\textbf{40.0} \pm 5.6$ \\ \bottomrule    
  \end{tabular}
  }
  \vspace{-12pt}
\end{table}
To compare our model with the baselines under the zero-shot setting, we mimic
this scenario by randomly splitting original category set into three parts.
For the MET task, we allocate 4 categories to each of the training, validation,
and test sets individually.
In the case of the MRE task, we assign 8, 7, and 7 categories to the training,
validation, and test sets respectively.
Ultimately, we conduct the experiments 3 times using different seeds and report
the mean and standard deviation of the performances. The hidden layers are
configured with a size of 768, while the hyper-parameter $\beta$, expert numbers
$K$, learning rate and batch size are set to 1.0, 8, 1e-5 and 16, respectively.
For both tasks, we set the training epochs to 20. All experiments are expedited
using NVIDIA GTX A6000 devices.

\subsection{Compared Methods}
We compare MG-VMoE with the text-based ZS-IE models and multimodal ones to
demonstrate its effectiveness.
For ZS-IE, the label-embedding-based prototype (Proto)
network~\cite{DBLP:conf/coling/MaCG16} is an effective baseline to perform the
tasks.
Besides, we select the description-based ZS-ET models including:
DBZFET~\cite{DBLP:conf/naacl/ObeidatFST19} and
NZFET~\cite{DBLP:conf/www/RenLZ20} as baselines which exploited attention
mechanisms to extract local features.
The memory augmented model MZET~\cite{DBLP:conf/coling/ZhangXLY20} was
introduced to transfer knowledge from observed types to unobserved ones.
For ZS-RE, \citet{chen-li-2021-zs} proposed ZS-BERT to encode entity and context
representations and connect them with semantic ones of relations.
\citet{DBLP:conf/acl/ZhaoZZZGWWPS23} exploited the fine-grained matching
mechanism to extract the effective entity and context features.
And \citet{DBLP:conf/coling/GongE24a} used prompt learning to model the
representations of both seen and unseen relations.
For ZS-MIE, we extend Proto with the pre-trained vision model as the multimodal
prototype network (MMProto)~\cite{DBLP:conf/aaai/WanZDHYP21}.
\citet{zhang-etal-2023-incorporating-object} proposed the multimodal entity
typing method (MOVCNet) to capture the semantic correlation between textual and
visual representations.
With the development of multimodal large language models, we select 13B version
of LLaVA~\cite{DBLP:conf/nips/LiuLWL23a} as the baseline to perform the ZS-MIE
tasks.

\subsection{Experimental Results}
\begin{figure}[t]
    \centering
    \includegraphics[width=\columnwidth]{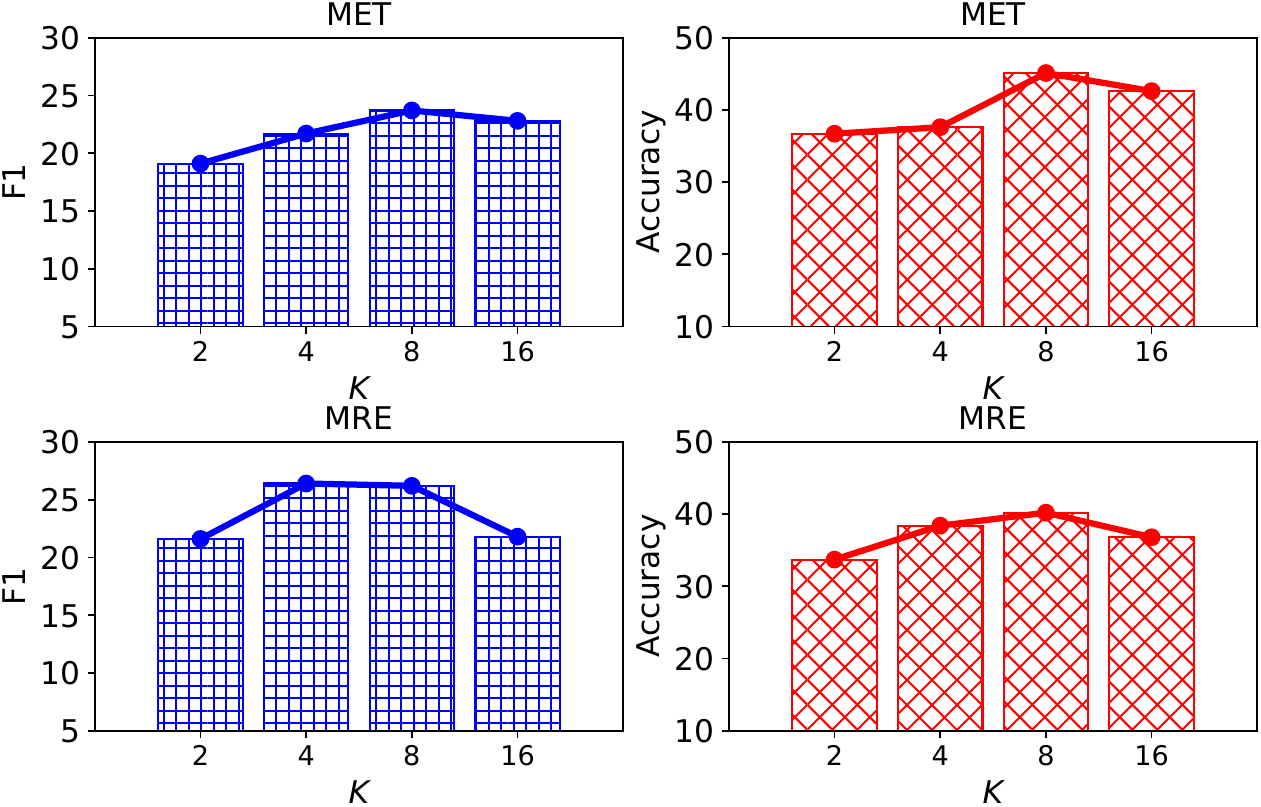}
    \setlength{\abovecaptionskip}{-0.2cm}
    \caption{
        The performance comparison of MG-VMoE with different expert module
        numbers.
    }
    \label{fig:expert_number}
    \vspace{-12pt}
\end{figure}
We evaluated MG-VMoE alongside baseline models on the MET and MRE benchmark datasets, and reported macro-averaged precision (P), recall (R), F1 scores, and accuracy, considering the varying sample sizes across different categories. The detailed experimental results are shown in \tbref{tab:main-results}.
Our model performed exceptionally well in most metrics on the MET dataset, achieving F1 and accuracy scores that exceeded the traditional baselines by 6.6\% and 15.8\%, respectively.
Among text-based methods, DBZFET consistently outperformed the Proto model in terms of F1 score, emphasizing the significance of the attention module in capturing local sentence representations pertinent to entity types.
We introduce the MG-VMoE model for fusing multimodal data to recognize novel types, consistently surpassing text-based models in performance, thereby demonstrating the superiority of our proposed approach.
But LLaVA achieved better results than our model and we analyze that LLaVA is based on LLaMA~\cite{DBLP:journals/corr/abs-2302-13971} which was pre-trained on the web contents of Wikipedia.
The MET dataset, sourced from the Wikinews website, includes entities that are also documented on Wikipedia.
Therefore, the results of LLaVA are higher than those of the MG-VMoE model.

On the MRE dataset, our model surpassed all baseline models, achieving an F1 score that was 3.9\% higher and an accuracy score that was 2.0\% higher compared to the baselines.
%
%
Furthermore, the multimodal-based MMProto exceeded text-based models in F1 score, highlighting the beneficial impact of visual information on ZS-MIE tasks.
And LLaVA fails to surpass the proposed model, as it was not pre-trained on the Twitter content comprising the MET dataset.
In conclusion, our model outperforms both text-based and multimodal-based baselines because of our novel fusion of text and image information using the VMoE network, combined with the creation of MG-VAT, which enhances multimodal representations and ultimately benefits MG-VMoE.

\subsection{Ablation Study}
The results demonstrate the significant role each component plays in determining
the model's overall performance.
To fully assess the effectiveness of the different modules introduced in MG-VMoE, an ablation study was performed, and its results are displayed in \tbref{tab:ablation-study}.
Significantly, the exclusion of the variational mixture of experts (VMoE) caused a substantial decline in performance, underscoring the importance of integrating aligned and informative multimodal representations. This decline occurs because the low-level features from both modalities possess complementary information, which collectively enhances the comprehension of the input data. By utilizing VMoE to fuse these features, the model is able to capture fine-grained semantic correlations between the modalities, resulting in enhanced performance.
\begin{figure}[t]
    \centering
    \includegraphics[width=\columnwidth]{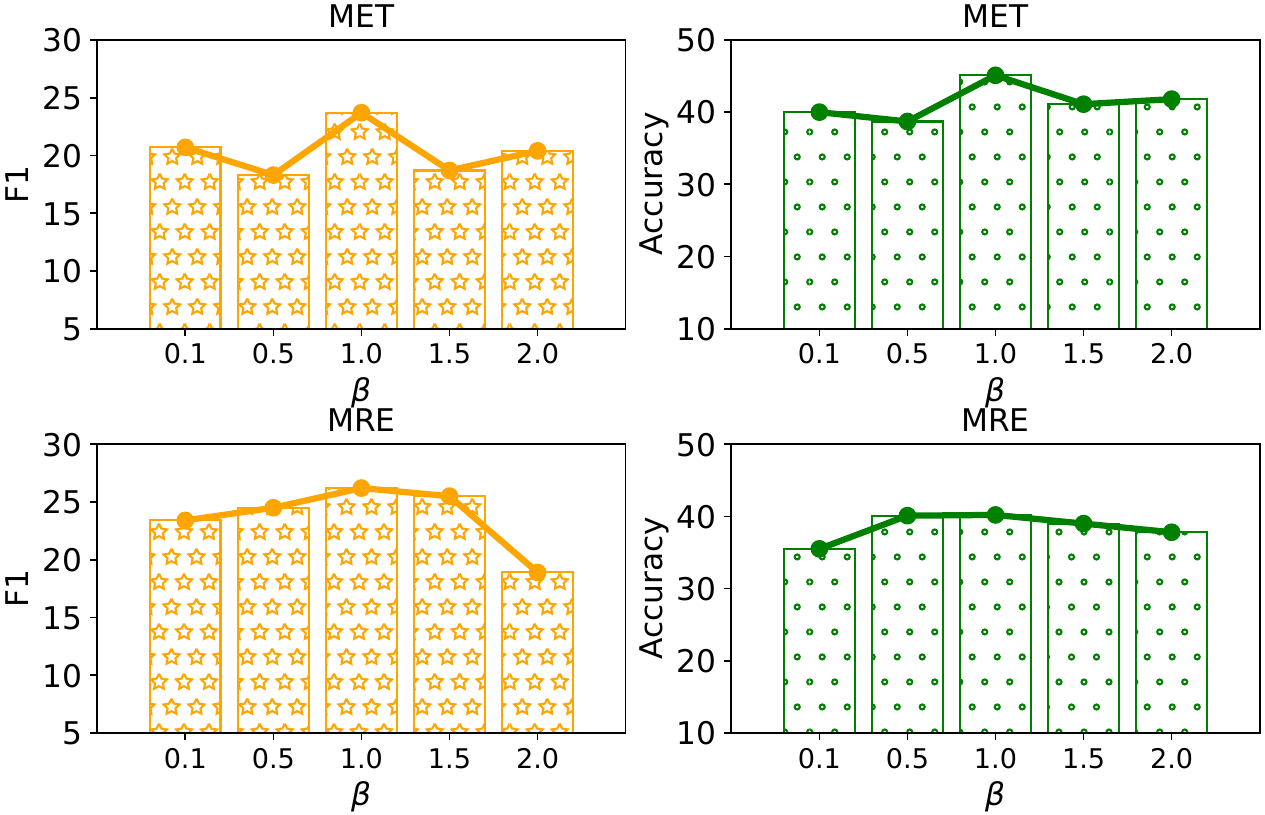}
    \setlength{\abovecaptionskip}{-0.2cm}
    \caption{
        The performance comparison of MG-VMoE with different hyper-parameter $\beta$ values.
    }
    \label{fig:beta_results}
    \vspace{-8pt}
\end{figure}
\begin{table}[t]
  \setlength{\abovecaptionskip}{0.1cm}
  \caption{
      The ablation study results of MG-VMoE on the MET and MRE benchmark datasets.
  }
  \label{tab:ablation-study}
  \begin{tabular}{l|cc|cc}
      \toprule
      \multirow{2}{*}{Model}   & \multicolumn{2}{c|}{MET} & \multicolumn{2}{c}{MRE} \\
                               & F1       & Accuracy     & F1       & Accuracy     \\ \midrule
      MG-VMoE                  & 23.7     & 45.1         & 27.3     & 40.0         \\
      ~~w/o VMoE               & 19.4     & 29.2         & 20.8     & 36.0         \\
      ~~w/o MG-VAT             & 18.2     & 38.9         & 21.5     & 32.5         \\ \bottomrule
  \end{tabular}
  \vspace{-12pt}
\end{table}
Additionally, we assess the impact of multimodal graph-based virtual adversarial training (MG-VAT) on the ultimate outcomes. Our findings indicate that MG-VAT plays a crucial role in boosting the model's final performance. By adopting an adversarial training strategy to refine multimodal representations at a fine-grained level, MG-VAT augments the model's capacity to precisely capture semantic connections among multimodal samples, leveraging graph-based information.
To summarize, the results obtained confirm that employing the VMoE network architecture alongside the MG-VAT strategy for modeling detailed multimodal representations can improve model performance.

\subsection{Influence of Experts Numbers}
The VMoE network consists of expert modules, and their quantity plays a crucial
role in the network's performance. To assess the impact of the number of
experts, we conducted experiments to compare the performance of MG-VMoE with
varying numbers of expert modules, as illustrated in \figref{fig:expert_number}.
The varying number of expert modules affects the MG-VMoE's performance on the
two benchmark datasets, with optimal results achieved when the model includes 8
experts.
A significant drop in performance occurs when reducing the number of experts in
the model, as the experts are individually trained to capture multimodal
representations of samples across various categories, and with fewer experts,
effective learning of these representations becomes impractical.
\begin{figure}[t]
    \centering
    \includegraphics[width=\columnwidth]{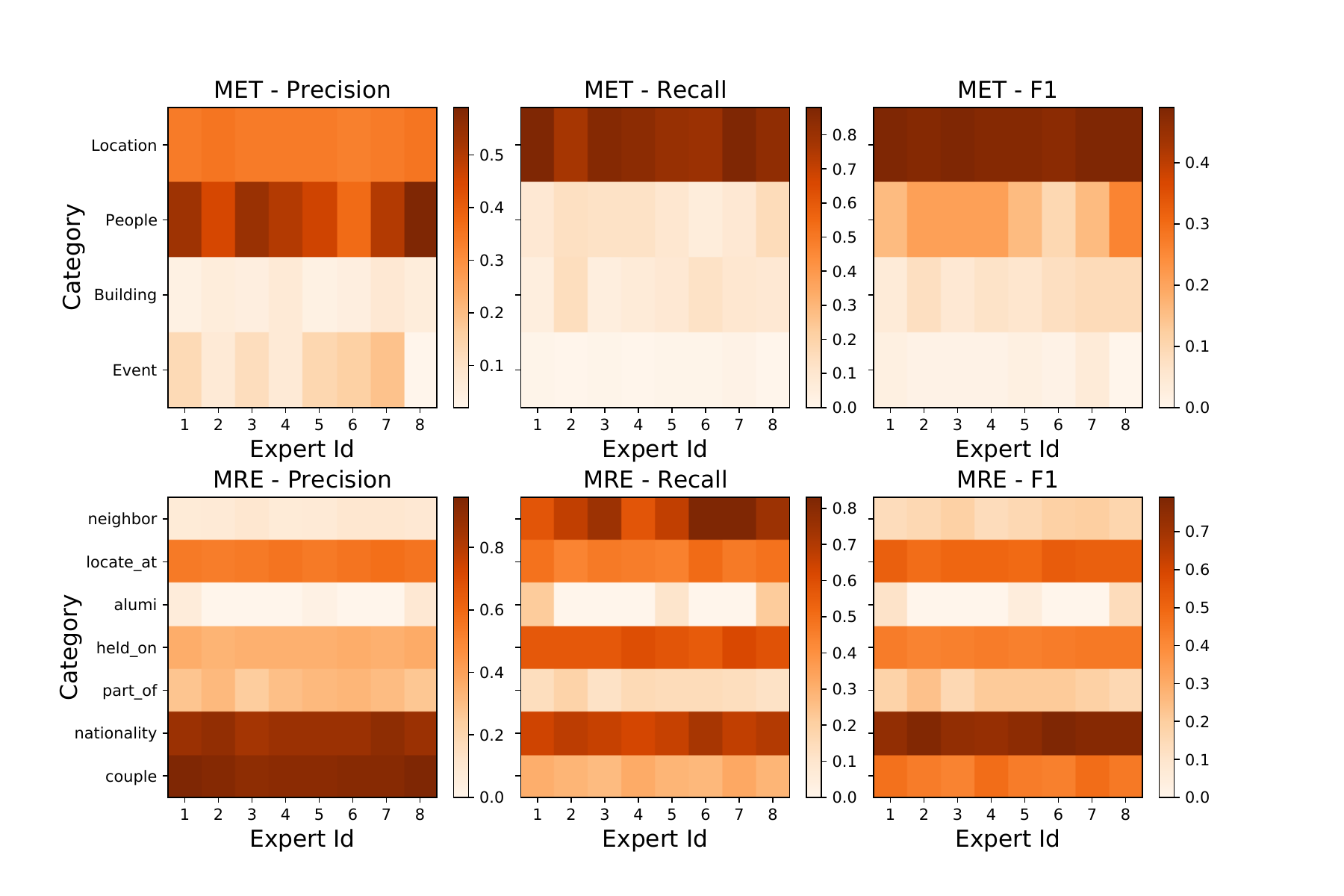}
    \setlength{\abovecaptionskip}{-0.2cm}
    \caption{
        The performance comparison of each expert module individually activated in MG-VMoE.
    }
    \label{fig:expert_ability}
    \vspace{-12pt}
\end{figure}
Increasing the number of experts in the model does not yield better results, as redundant experts may lead to overfitting on the training samples.
Furthermore, the varying outcomes of MG-VMoE with different numbers of experts demonstrate the VMoE network's effectiveness.

\subsection{Influence of Hyper-parameter $\beta$}
During the training process, we uniformly train the model by summing the loss functions of VMoE and MG-VAT with the ranking loss, weighted by a hyper-parameter. To assess the impact of the hyper-parameter $\beta$, we perform parameter sensitive experiments as shown in \figref{fig:beta_results}.
The MG-VMoE's performance on the two benchmark datasets is influenced by various $\beta$ values.
As $\beta$ value increases, the model focuses more on VMoE and MG-VAT losses, resulting in a significant drop in F1 scores. This is because the model fails to obtain a meaningful signal from the ranking loss and is unable to learn generalized representations for diverse categories.
When $\beta$ value is reduced, the model struggles to obtain sufficient signals from VMoE and MG-VAT losses, leading to a decline in both F1 and accuracy scores. This is due to the model's inability to effectively leverage VMoE and MG-VAT, resulting in its failure to learn aligned and informative multimodal representations for different categories.
Therefore, we achieve optimal results by balancing the ranking loss with other losses, setting the value of $\beta$ to 1.0.

\subsection{Experts Abilities Analysis}
During our experiments, MG-VMoE comprises 8 expert modules that are activated using sparse weights for forward propagation.
To evaluate the capabilities of various experts, we examine the performance of each expert module independently activated within MG-VMoE, as illustrated in \figref{fig:expert_ability}.
While the experts exhibit comparable overall performance, there are slight discrepancies in their prediction outcomes, particularly in terms of precision, recall, and F1 scores.
Each expert acquires diverse multimodal representations for samples belonging to different categories. On the MET dataset, Expert \#7 outperforms other experts in terms of results for the "Event" type, but falls behind Expert \#8 for the "People" type.
Furthermore, the recall scores of experts on the MRE dataset vary significantly.
Considering the diverse capabilities of experts demonstrated in \figref{fig:expert_ability}, as well as the inferior performance of the model with fewer experts shown in \figref{fig:expert_number}, having an optimal number of experts is essential for accurately modeling multimodal representations and enhancing the performance of MG-VMoE.

\subsection{Visualization Analysis}
\begin{figure}[t]
    \centering
    \includegraphics[width=\columnwidth]{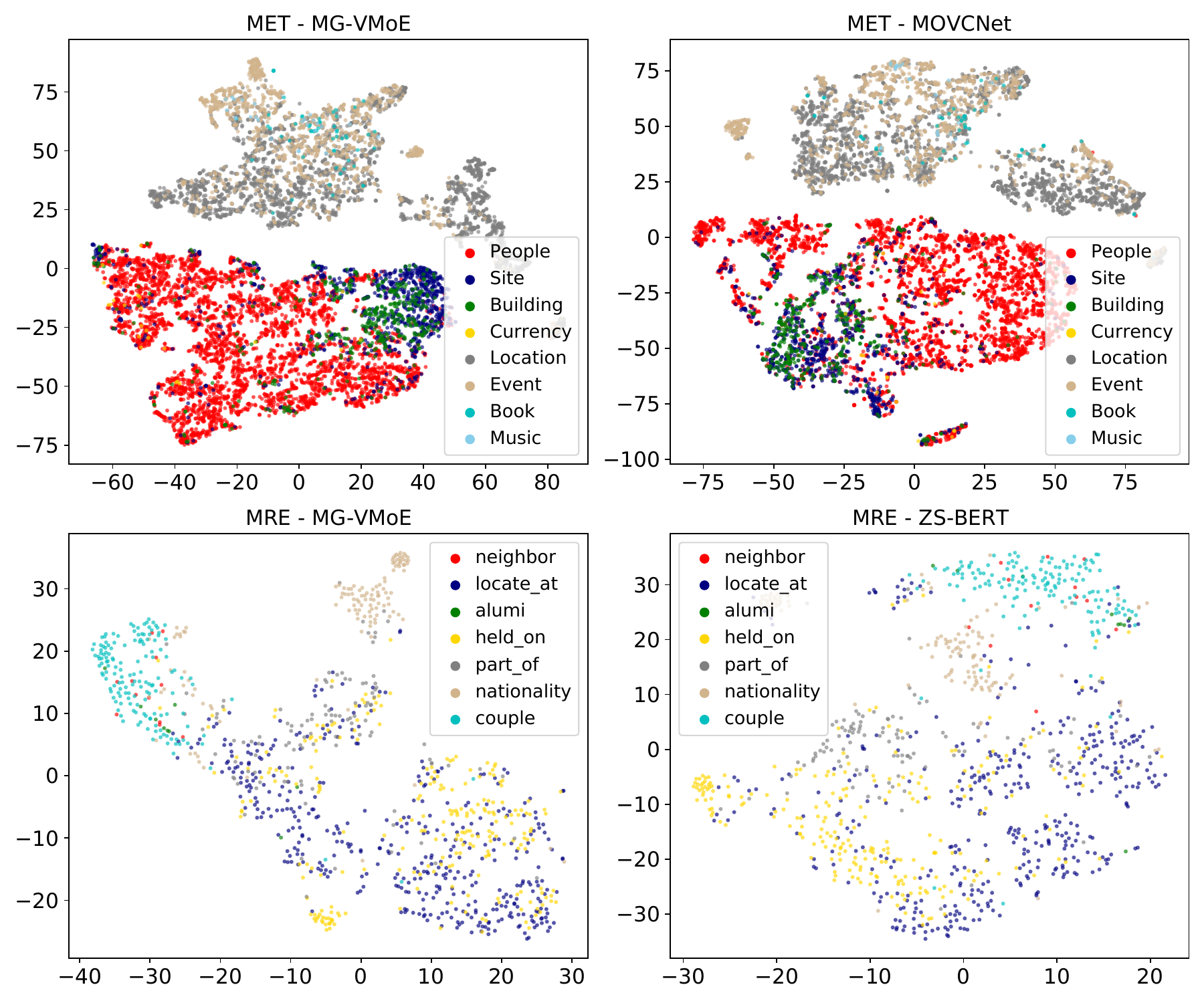}
    \setlength{\abovecaptionskip}{-0.2cm}
    \caption{
        Visualization of the t-SNE results exhibiting the multimodal representations of samples with particular categories, which were extracted individually from MG-VMoE and baseline models.
    }
    \label{fig:feature_analysis}
    \vspace{-12pt}
\end{figure}
In order to evaluate the effectiveness of multimodal representations for ZS-MIE,
we visualize the features learned from MG-VMoE and baseline models in
\figref{fig:feature_analysis}. We choose samples from specific categories within
the test sets and utilize both MG-VMoE and baseline models to obtain their
respective representations. Subsequently, we employ t-SNE~\cite{2008Visualizing}
to reduce the dimensionality of these output representations to two dimensions.
The results show that the representations of MG-VMoE are more tightly grouped
within each category, suggesting that MG-VMoE is more adept at distinguishing
subtle variations among samples within the same category.
For instance, the representations generated by MG-VMoE are more clustered
compared to those produced by ZS-BERT on the MRE dataset. Furthermore, on the
MET dataset, the representations of various categories from MG-VMoE exhibit a
tighter grouping within each category.
This can be explained by the abundant semantic content in multimodal data and
the effectiveness of MG-VAT in capturing semantic relationships between samples.
By integrating multimodal information using MG-VAT, the model acquires robust
features, leading to improved performance in ZS-MIE tasks.

\section{Conclusion}
This paper investigates zero-shot multimodal information extraction (ZS-MIE)
tasks, and mainly aims to address the coarse-grained multimodal representation
learning limitation. To overcome this limitation, we introduce the multimodal
graph-based variational mixture of experts (MG-VMoE) network tailored for ZS-MIE
tasks. The MG-VMoE network builds upon fine-grained multimodal representation
learning, incorporating both the variational mixture of experts (VMoE) and
multimodal graph-based virtual adversarial training. Serving as the core, the
VMoE network utilizes sparse weights to activate expert modules, where each
expert functions as a variational information bottleneck (VIB) for extracting
informative and aligned textual and visual representations. Meanwhile, the
multimodal graph-based virtual adversarial training is employed to capture
semantic correlations between multimodal samples and enhance the clustering
tightness of samples within the same category. Experimental results demonstrate
the generalization ability of MG-VMoE compared to baseline methods on ZS-MIE
tasks.

\begin{acks}
This research is supported by the National Natural Science Foundation of
China (No. 62272250, 62302243), the Natural Science Foundation of Tianjin, China
(No. 22JCJQJC00150, 23JCYBJC01230).
\end{acks}

\bibliographystyle{ACM-Reference-Format}
\balance
\bibliography{refs}

\end{document}